\documentclass{elsarticle}
\usepackage{amsmath}
\usepackage{amsfonts}   
\usepackage{graphicx}
\usepackage{color}

\newcommand{\ui}{\boldsymbol{\hat \i}}
\newcommand{\uj}{\boldsymbol{\hat \j}}
\newcommand{\uk}{\boldsymbol{\hat k}}
\newcommand{\newvec}[1]{\boldsymbol{#1}}

\begin{document}

\begin{abstract}
A Hamiltonian approach is presented to study the two dimensional motion of 
damped electric charges in time dependent electromagnetic fields. The 
classical and the corresponding quantum mechanical problems are solved for 
particular cases using canonical transformations applied to Hamiltonians for a
particle with variable mass.
The Green's function is constructed and, from it, the motion of a Gaussian 
wave packet is studied in detail.
\end{abstract}

\title{Quantum and classical dissipation of charged particles}
\author{V.G. Ibarra-Sierra$^1$,
        A. Anzaldo-Meneses$^2$,
        J.L. Cardoso$^2$,
        H. Hern\'andez-Salda\~na$^2$,
        A. Kunold$^{2}$
        and
        J. A. E. Roa-Neri$^2$}
\address{
   $^1$ Departamento de F\'isica, Universidad Aut\'onoma Metropolitana
   at Iztapalapa, Av. San Rafael Atlixco 186, Col. Vicentina,
   09340 M\'exico D.F., Mexico
\\
   $^2$\'Area de F\'isica Te\'orica y Materia Condensada,
   Universidad Aut\'onoma Metropolitana at Azcapotzalco,
   Av. San Pablo 180, Col. Reynosa-Tamaulipas, Azcapotzalco,
   02200 M\'exico D.F., M\'exico }

\maketitle

\section{Introduction}

The motion of particles in vacuum and diverse media with dissipation has been 
studied in classical and quantum physics since long time. An important class 
of such problems are those of free electric charge carriers in a material 
under external time dependent electromagnetic fields. Of particular interest 
is the dissipation of energy through the interaction of charged carriers with 
the lattice ions (phonons) of the material, the carrier to carrier interaction
through a Coulombian potential  and, eventually, through radiation.

In classical systems damping is often described by including a velocity 
dependent drag term in Newton's second law. However, the inclusion of dissipation 
phenomena in quantum mechanics requires special care since its building 
blocks, time independent Hamiltonians, lead to energy conservation. This 
shortcoming is remedied in the heat bath approach \cite{caldeira1, caldeira2} 
by coupling the single particle Hamiltonian with an infinite degrees of 
freedom system, e.g., an infinite collection of harmonic oscillators, to which 
the energy of the single particle is transferred. Even though the energy is 
conserved given that the single particle, heat bath and coupling Hamiltonians 
are time-independent it is difficult to handle calculations with the many 
degrees of freedom of the heat bath\cite{heinzpeter}. The dynamics of an open 
quantum system\cite{lindblad} is often formulated in terms of a master 
equation for the density matrix, that allows to work only with the single 
particle degrees of freedom by adding extra terms to the Von Neumann equation. 
Notwithstanding, the change over time of the open quantum system, in general, can not 
be presented in terms of a unitary time evolution\cite{heinzpeter}.
{ Other approaches to quantum dissipation
include the use of effective Schr\"odinger equations
\cite{razavy2,pimpale:2739} and functional integration \cite{dittrich,kochan:022112}.}

In this work, we treat the problem of energy dissipation by means of a single 
charged particle \emph{time dependent Hamiltonian} \cite{bateman, caldirola, 
kanai}. In contrast to time independent Hamiltonians, in the time dependent 
ones the energy is no longer a conserved quantity and therefore they allow for 
the possibility of energy loss. In particular, we study
{ the Hamiltonian
of a charged particle with minimal coupling
under a time dependent electromagnetic field
with a variable mass term that accounts for energy loss. 
Eventhough it has been shown that the use of minimal coupling
procedure to switch on electromagnetic interactions in phenomenological
quantum equations of damped motion leads to incorrect equations in
the classical limit\cite{pimpale:2739} we show that the standard Schr\"odinger
equation with minimal coupling and a variable mass produce correct results
for the stationary state of the particles motion and allow for the
modeling of the transient state by means of the time dependence
of the mass.}

In classical mechanics friction is usually analyzed introducing an opposing 
velocity-proportional force. The equation of motion of the particle can be 
usually built without difficulties from Newton's second law of motion. For a 
one dimensional particle with mass $m$ subject to a potential $U$ one has
\begin{equation}\label{ecmotion}
m \ddot x+\frac{m}{\tau}\dot x+\frac{\partial U}{\partial x}=0,
\end{equation}
where $x$ is the position of the particle and $\tau$ is the collision
time.

A deeper dynamical analysis is reached when the Hamiltonian formalism is 
applied. In the special case of one dimensional movement described by Eq. 
(\ref{ecmotion}) the dynamics of a particle may be expressed by the 
Kanai-Caldirola (KC) Hamiltonian\cite{bateman, caldirola, kanai} 
\begin{equation}\label{kancal}
H=\frac{p^2}{2m}{\rm e}^{-t/\tau}+U\left(x\right){\rm e}^{t/\tau}.
\end{equation}
This Hamiltonian even allows for analytical treatment in some simple
quantum mechanical systems as a free particle\cite{cavalcanti:6807}
($U=0$) and the harmonic oscillator\cite{kener:371, bopp:699, stevens:1072, 
englert:1,hasse:2005,edwards:153}.

A great deal of effort has been focused on the modeling of dissipation 
phenomena for a charged particle through time dependent Hamiltonians 
\cite{schuch:6571, razavy}. However, obtaining  a Hamiltonian for a 
dissipative charged particle under electric and magnetic fields is not as 
straightforward as for the KC Hamiltonian (\ref{kancal}). The
 assumption of a damping force proportional to the velocity does not 
lead to a Hamiltonian formulation, i.e., the Newton's equations of motion
\begin{eqnarray}
m \ddot x+\frac{m}{\tau}\dot x+q B\dot y-q E_x &=& 0\label{naivex},\\
m \ddot y+\frac{m}{\tau}\dot y-q B\dot x-q E_y &=& 0\label{naivey},
\end{eqnarray}
of a particle in perpendicular electric and magnetic fields $E_x\ui+E_y\uj$ 
and $B\uk$ respectively can not be obtained from a Hamiltonian approach.
From here on we call this the Newtonian model.

Nevertheless, as we shall demonstrate below, it is possible to model 
dissipation by introducing a time-dependent mass in the Hamiltonian for a  
charged particle
\begin{equation}
{H}= \frac{1}{2m\left(t\right)}\left(\newvec{p}
    -q\newvec{A} \right)^2+q \phi +V.\label{hammt}
\end{equation}

The aim of this work is to study the dynamics of a damped charged particle in 
the presence of time dependent perpendicular electric and magnetic fields by 
means of a time dependent Hamiltonian. We obtain the general solutions for the 
equations of motion for the classical, as well as for the quantum problem, via the 
reduction of the Hamiltonian to zero  by means of a series of linear canonical 
transformations in the classical case and corresponding unitary 
transformations in the quantum mechanical one. Here it is important to stress 
that, in general, in a large kind of dynamical systems the number of constants 
of motions is not enough to reduce the Hamiltonian to zero\cite{arnold}. 
In this work, it 
is assumed that the Hamiltonian is at most quadratic in the canonical 
coordinates, so that $\newvec{A}$ is at most linear in the generalized positions, but 
the scalar potentials can be quadratic.

The well known classical and quantum dynamics for a constant or a variable 
mass charged-particle in constant perpendicular electric and magnetic fields  
are recovered from our analysis. 

This paper is organized as follows. In Sec. \ref{varmas} we review the role of 
time dependent masses in the Hamiltonian of charged particles interacting with 
electromagnetic fields. In Sec. \ref{secclas} we address the solution of the
classical Hamiltonian via canonical transformations. The quantum mechanical 
problem is introduced in Sec. \ref{quantization}. Unitary transformations are 
applied to reduce the quantum mechanical Hamiltonian in Subsec. 
\ref{unitoperator}. With the resulting time evolution unitary operator, the 
Green's function is derived in Subsec. \ref{green}. As an example we study 
the dynamics of a Gaussian wave packet under the action of the Hamiltonian 
solved in this paper in Subsec. \ref{gaussian}. We conclude in Sec. 
\ref{conclusions} with a summary of the results.


\section{Hamiltonian with a variable mass.}\label{varmas}

To study the above physical problems a geometric setting is adopted. Let the 
kinetic energy $T$ be given by a smoothly varying family of Riemannian metrics
$\langle \dot{\newvec{r}},g\dot{\newvec{r}}\rangle=\sum_{ij} g_{ij}(x,t) 
\dot{x}_i \dot{x}_j$, parametrized by time $t$ on a $n$-dimensional 
manifold. The Lagrangian is then
{ 
\begin{equation}
L= T -q\phi+q\newvec{A}\cdot \dot{\newvec{r}},
\end{equation}
in terms of the vector potential $\newvec{A}$ and the scalar potential $\phi$.
}
The Hamiltonian is given
by the Legendre transformation of the generalized velocities,
{
\begin{eqnarray}
H &=& \sum_ip_i\dot{x}_i-L=T+q\phi- q\newvec{A}\cdot \dot{\newvec{r}},\\
 p_j &=& \frac{\partial L}{\partial \dot{x}_j}= \sum_kg_{jk}\dot{x}_k+ qA_j,
\end{eqnarray}
}
and leads to a kinetic energy given in terms of the momenta as 
\begin{equation}
T=\frac{1}{2}\left\langle \newvec{p}- q\newvec{A},g^{-1} \left(\newvec{p}
- q\newvec{A} \right) \right\rangle
= \frac{1}{2}\sum_{ij} (g^{-1})_{ij}
\left(p_i- qA_i \right) \left(p_j- qA_j \right).
\end{equation} 
In this approach it is assumed then, that the media acts on the particle by 
means of an alteration of the metric corresponding to replace the 
constant mass of the particle by a {\em time dependent} effective mass. Only the 
flat diagonal case $g_{ij}=\delta_{ij}m(t)$, with a time dependent mass, shall 
be studied in here. However, more general metrics could be introduced in this 
manner, for example to include space inhomogeneities\cite{ans}, but they shall not be 
considered in this work. 

Let us here start with the classical Hamiltonian for a charged particle
{
\begin{equation}
{H}= \frac{1}{2m}  \left(\newvec{p} -q  \newvec{A} \right)^2 + q \phi, 
\label{ham}
\end{equation}
}
with a time dependent mass $m$.
The equations of motion obtained from (\ref{ham}) are
{
\begin{eqnarray}
\dot{x}_i = \frac{\partial{H}}{\partial p_i} &=&
   \frac{p_i}{m}- \frac{qA_i}{m},\\
\dot{p}_i = - \frac{\partial{H}}{\partial x_i} &=&
 -\frac{q}{m }\sum_{j}  \frac{\partial A_j}{\partial x_i}\left( p_j-qA_j\right)
  - q\frac{\partial\phi }{\partial x_i}, 
\end{eqnarray}
}
written as the Newton's second law 
they take the following form
{
\begin{equation}
\frac{d}{dt} (m\dot{\newvec{r}})=q(\newvec{E} + \dot{\newvec{r}}
\times \newvec{B}).\label{New}
\end{equation}
} 
with $\newvec{B}= \newvec{\nabla}\times \newvec{A}$ and $\newvec{E}= 
-\newvec{\nabla}\phi -\partial_t\newvec{A}$. It must be emphasized that this equation
is obtained from a Hamiltonian variational principle.

In order to illustrate how to model dissipation through a time dependent mass
let us consider a charged particle in uniform perpendicular magnetic and
electric fields
\begin{eqnarray}
\newvec{B} &=& B\uk, \\
\newvec{E} &=& E_x\ui+E_y \uj.
\end{eqnarray}
Separating the two components of Eq. (\ref{New}) we obtain the
following equations of motion for the particle
{
\begin{eqnarray} 
m\ddot{x}+\dot m\dot{x}+m\omega\dot{y}-qE_x &=& 0,\label{newoldx}\\
m\ddot{y}+\dot m\dot{y}-m\omega\dot{x}-qE_y &=& 0,\label{newoldy}
\end{eqnarray}
}
where
\begin{equation}\label{cyclotron}
\omega=\frac{qB}{m},
\end{equation}
is, in general, time dependent. Notice that for an electron ($q=-e$) for
constant magnetic field and mass $\left\vert\omega\right\vert=\omega_c=eB/m$
is the cyclotron frequency.
In stationary state { ($m\ddot{x}=m\ddot{y}=0$)}, the solution for
these equations is
\begin{eqnarray}
\dot x &=& \frac{q}{\dot m}
      \frac{E_x+(q B/\dot m)E_y}{1+(q^2B^2/\dot m^2)},\label{gstatx}\\
\dot y &=& \frac{q}{\dot m}
     \frac{E_y-(q B/\dot m)E_x}{1+(q^2B^2/\dot m^2)}.\label{gstaty}
\end{eqnarray}
In order to test the time dependent mass model
equations, specially the ones that describe the stationary state,
let us try two different time dependent mass models.
First we consider a KC-like mass \cite{schuch:6571, razavy}
\begin{equation}
m=m_0{\rm e}^{t /\tau},\label{calkanmass}
\end{equation}
where, for example, $m_0$ and $\tau$ may be related to the effective mass
and collision time in a semiconductor with mobility $\mu_e=nq^2\tau/m_0$
and charge carrier density $n$.
Dislike the Newtonian model, in this case, the time dependent mass model yields  vanishing
velocity components even in the presence of an electric field. Well known results,
as the magneto conductivity tensor in semiconductors \cite{chakraborty},
are contradicted by this calculation.
  
As a second example let us consider the following convenient choice
of the mass' time dependence
\begin{equation}
m=m_0\left(\frac{t}{\tau}+k\right),\label{masdep}
\end{equation}
{
where $k$ is a dimensionless positive parameter.
We shall call this the linear time dependent mass model (LTDMM, for "short'').
Eq. (\ref{New}) can be conveniently recast as
\begin{equation}
m\ddot{\newvec{r}}=q(\newvec{E} + \dot{\newvec{r}}
\times \newvec{B})-\dot m\dot{\newvec{r}}.\label{secondlawnewt}
\end{equation}
The two first terms in the right hand side of this equation correspond to the Lorentz 
force whereas the last term accounts for damping. Indeed, for the LTDMM
\begin{equation}
\dot m\dot{\newvec{r}}=\frac{m_0}{\tau}\dot{\newvec{r}}.
\end{equation}
Here it is important to keep in mind that, despite the resemblance between
Eq. (\ref{secondlawnewt}) and the Newtonian model in Eqs. (\ref{naivex}) and (\ref{naivey}),
in the former the mass is time dependent. Despite this difference, the stationary state
for both models is the same. 
For the LTDMM  the stationary state solution for the
velocity components is in fact
}
\begin{eqnarray}
\dot x &=& \frac{q\tau}{m_0}
      \frac{E_x+\omega_0\tau E_y}{1+\omega_0^2\tau^2},\label{lstatx}\\
\dot y &=& \frac{q\tau}{m_0}
      \frac{E_y-\omega_0\tau E_x}{1+\omega_0^2\tau^2}\label{lstaty}
\end{eqnarray}
where $\omega_0=\omega\left(t=0\right)$.
Thus, our LTDMM approach and the Newtonian model given by
Eqs. (\ref{naivex}) and (\ref{naivey}) yield the same non vanishing stationary state
solution even-though their transient states might be slightly different.

{ However similar to the Newtonian approach, the LTDMM is only physically
meaningful for
$t > -k$ given that for $t \le -k $ the mass becomes zero or even negative.
One can overcome this limitation by proposing more complex models as
\begin{equation}
m\left(t\right)=m_0 \ln\left(1+{\rm e}^{ t/\tau }\right),
\end{equation}
that yield positive non vanishing masses for all finite times and, regardless of its
complexity, the same stationary state as the Newtonian and LTDMM models.}
Notice that this model interpolates between the KC model for $t\rightarrow -\infty$ and
the LTDMM for $t\rightarrow \infty$.

To provide with a numerical example we have chosen a charged particle, e.g.,
an electron, in a GaAs sample with mobility
$\mu_e=nq^2\tau/m_0=148 m^2/Vs$, that yields a collision time $\tau=56 ps$. The
effective mass and charge will be set to $m_0=0.067 m_e$ and $q=-e$, respectively,
with $e$ the electric charge of the electron. The magnetic and
electric fields are $B=40mT$  and $\boldsymbol{E}=100 V/m\uj$.
The initial position and velocity of the particle are set to the origin
and to
$\boldsymbol{\dot r}=\dot x\left(0\right)\ui+\dot y\left(0\right)\uj=3.7 Km/s \uj$,
respectively.

\begin{figure}
\includegraphics[width=8 cm]{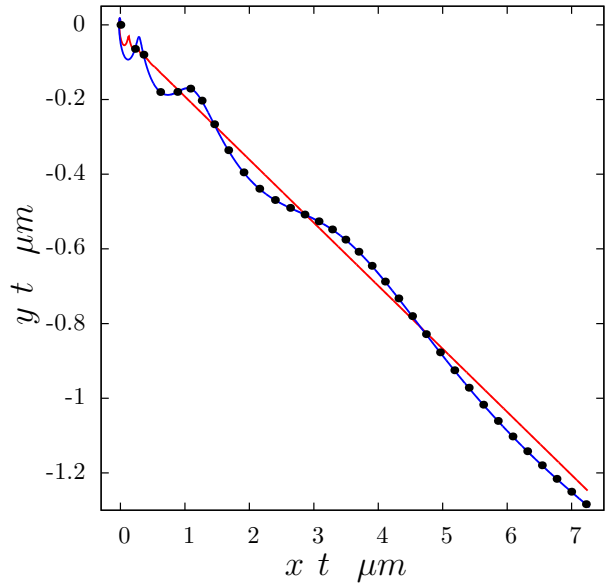}
\caption{
(color online). Trajectory of a charged particle for the Newtonian
model (red solid line) given by Eqs.(\ref{naivex}) and (\ref{naivey}), the linear time 
dependent mass model (LTDMM) defined in Eq. (\ref{masdep}) (blue solid line) and
governed by  Eqs.(\ref{newoldx}) and (\ref{newoldy}) and the center of the quantum
mechanical Gaussian wave packet given by Eqs. (\ref{lambdaxR})-\ref{lambday0R}) (black dots).}
\label{figure1}
\end{figure}

Fig. \ref{figure1} shows a comparison between the parametric plots
of $\boldsymbol{r}\left(t\right)=x\left(t\right)\ui+y\left(t\right)\uj$
for the Newtonian model (red) and the LTDMM (blue). We observe
that even-though both models present different trajectories for the transient
state in $t\rightarrow \infty$ they have the same overall behavior.

\begin{figure}
\includegraphics[width=8 cm]{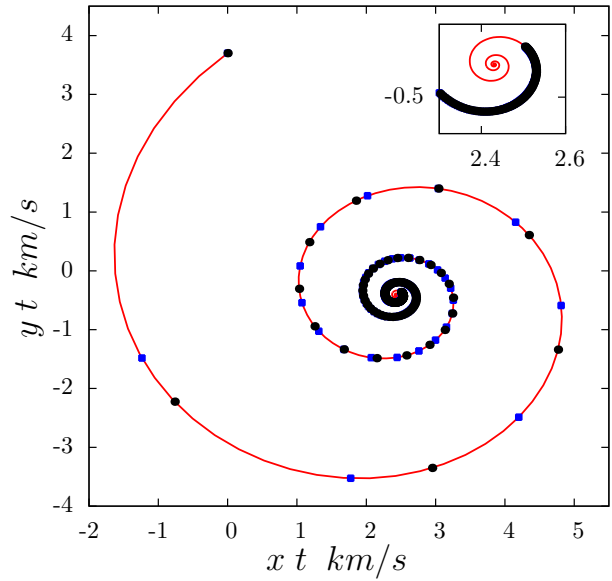}
\caption{
(color online). Parametric plot of the velocity components
for a charged particle for the Newtonian model (red solid line)
and the LTDMM for the classical case (blue dots) and the
center of a quantum mechanical Gaussian wave packet (black points) as
calculated in Subsec. \ref{gaussian}.}
\label{figure2}
\end{figure}

\begin{figure}
\includegraphics[width=8 cm]{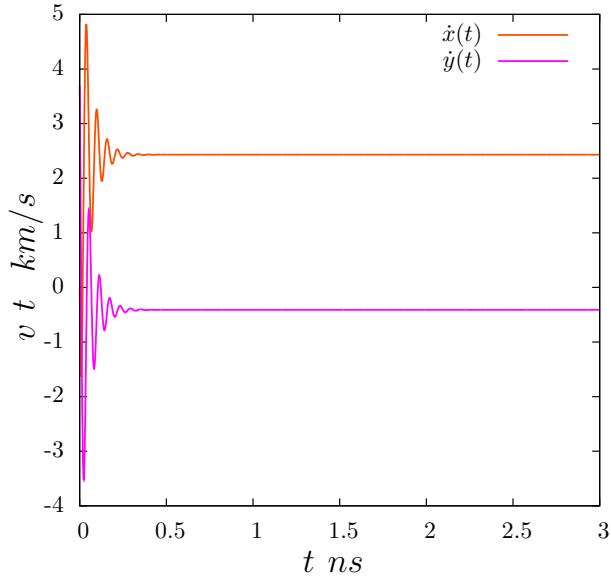}
\caption{
(color online). 
Velocity components, $\dot x$ (blue solid line) and
$\dot y$ (red solid line), as functions of time for the Newtonian
model.}
\label{figure3}
\end{figure}

\begin{figure}
\includegraphics[width=8 cm]{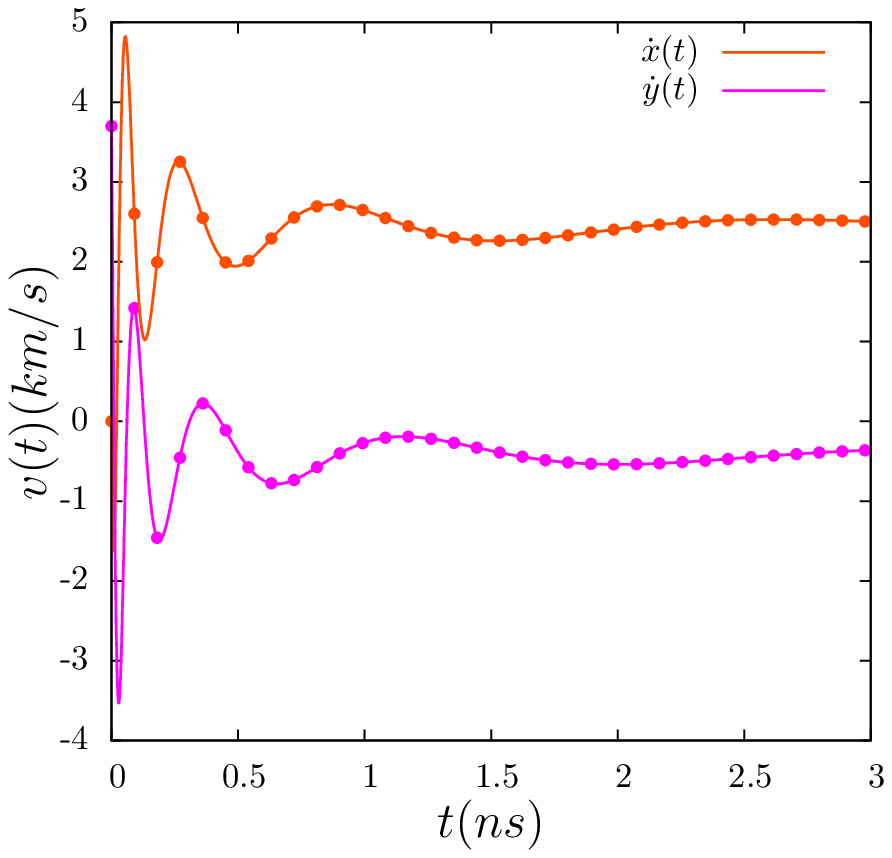}
\caption{
(color online). 
Velocity components, $\dot x$ and $\dot y$, as functions of time for the LTDMM
and for the center of the quantum
mechanical Gaussian wave packet $\dot{\zeta}_x^R$ and $\dot{\zeta}_y^R$ (points) as well.}
\label{figure4}
\end{figure}

In Fig. \ref{figure2} we can see a parametric plot
of the velocity vector
$\boldsymbol{\dot r}\left(t\right)=\dot x\left(t\right)\ui+\dot y\left(t\right)\uj$,
for the LTDMM (blue dots)
given by (\ref{masdep}) and the Newtonian model (red solid line). Surprisingly both
models plots are clearly over the same curve. Nevertheless we can not say
that both examples behave exactly the same since the Newtonian model reaches
the terminal velocity faster than the LTDMM.
This is shown in Figs. \ref{figure3} and \ref{figure4} where we can
observe $\dot x$ and $\dot y$ plots for both models.
We appreciate that the Newtonian model saturates after $t=1$ ns
meanwhile the LTDMM saturates after $t= 2.5$ ns. 

Since the LTDMM yields similar results as the Newtonian one, and both reach the same 
stationary state, we shall use it through out the rest of the work for the numerical examples.
Notwithstanding, all the calculations in next sections 
do not rely on a specific mass model.
 
\section{The classical problem: canonical transformations}\label{secclas}

A possible procedure to solve analytically the previously depicted problem
is to perform a set of canonical transformations \cite{lewis1}.
Equivalently, also the proposal of a function of the canonical 
coordinates at most quadratic in the momenta, has been successful in similar 
problems \cite{lewis2}.
We chose this approach for the classical problem in order to establish 
a connection between the canonical and the unitary quantum transformation.

The reduction of the Hamiltonian (\ref{hammt}) is accomplished by applying canonical 
transformations of a certain sub-group of the affine group, namely, 
translations, dilatations, shears, and rotations in phase space $\xi=(x,p)^t$,  
\begin{equation}
\xi \mapsto \mathbf{M}\xi + \mu,
\end{equation}
with time dependent  vector $\mu$, and time dependent non-singular symplectic 
matrix $\mathbf{M}$.

The study of a charged particle's motion under homogeneous electric and magnetic time
dependent fields, is of utmost importance in the experimental and theoretical analysis 
of solid state devices.  
The building block of any theory explaining the integer and fractional quantum 
Hall effects \cite{klitzing:494, tsui:1559}, 
Shuvnikov-de Haas oscillations \cite{fang:174},
microwave induced resistance oscillations \cite{zudov:201311}, Hall induced 
resistance oscillations, amongst others, is the 2D electron in crossed electromagnetic 
fields.
Therefore,  we shall consider a 2D charge particle under perpendicular magnetic field 
\begin{equation} \label{magneticfield} 
\newvec{B}=B\uk,
\end{equation} 
with a vector potential given by
\begin{equation} 
\newvec{A}= -\frac{B}{2}y \ui + \frac{B}{2}x \uj. \label{potential:vector}
\end{equation}  
The  in-plane electric field is
\begin{equation} \label{electricfield}
\newvec{E}= \left(\frac{\dot{B}}{2}y+E_x\right)\ui
-\left(\frac{\dot{B}}{2}x -E_y \right) \uj .
\end{equation} 
 with a scalar potential 
\begin{equation}
\phi = -E_x x - E_y y. \label{potential:scalar}
\end{equation}
Here $B$, $E_x$ and $E_y$ are functions only of time. 

For the sake of simplicity and without any loss of generality, we have considered the 
simplest gauge transformation to write down the scalar and the vector 
potentials.

The resulting quadratic time dependent Hamiltonian for the mentioned fields is
\begin{multline}
H= \frac{1}{2m}\left(p_x^2+p_y^2\right) 
 +\frac{1}{8} \left( m \omega^2 +\kappa \right)  \left(x^2+ y^2 \right)\\
- \frac{\omega}{2} 
(x p_y- y p_x) 
-qE_x x  -qE_y y  , \label{ham2}
\end{multline}
{ where $\omega$ is in general
a time dependent parameter given by (\ref{cyclotron}).
In order to generalize the problem we have added a confining potential
$V=\kappa\left(t\right)( x^2+ y^2)/8$.
Since $z$ is cyclic, the momentum associated with it has been dropped but not forgotten.
Here all coefficients are given smooth functions of time. }

Our aim now is to 
reduce the Hamiltonian (\ref{ham2}) to zero using canonical transformations (see for 
example \cite{goldstein}) of a certain sub-group of the affine group. 
The procedure can be summarized as follows:
1) The third term in $H$, corresponding to the 
coupling $z$ component of the angular momentum, can be
eliminated by a rotation leaving the first two terms invariant.
The result is a Hamiltonian for two uncoupled one 
dimensional harmonic oscillators with variable masses and frequencies, as 
those considered in the literature;
2) a time dependent translation is 
performed to eliminate the linear contributions leading to an harmonic 
oscillator Hamiltonian with time dependent coefficients;
3) and, finally a dilatation and two shears are applied to reduce the Hamiltonian to 
zero; hence,  
the final generalized momenta and positions are simultaneously constants of the motion 
and the original initial conditions, $(p_0,q_0)$, of our problem. 

For the first step in our program we require 
the generating function of a rotation $R$ for a finite angle $\theta(t)$ given by 
\begin{eqnarray} \label{F1}
F_1=q^tR^tp_1
=xp_{x_1} \cos \theta + yp_{x_1} \sin \theta - xp_{y_1} \sin \theta + yp_{y_1} \cos \theta ,
\end{eqnarray}
with the column vectors $q=(x,y)^t$ and $p_1=(p_{x_1},p_{y_1})^t$, being 
$p_1=Rp$ and $q_1=Rq$ the rotated coordinates. By means of this generating 
function we obtain the following transformation rules
\begin{eqnarray}
x_1 &=& \frac{\partial F_1}{\partial p_{x_1}} =   x \cos \theta  + y \sin \theta,
\label{F1:configx} \\
y_1 &=& \frac{\partial F_1}{\partial p_{y_1}} =  -x \sin \theta  + y \cos \theta,
\label{F1:configy} 
\end{eqnarray}
and
\begin{eqnarray}
p_x &=& \frac{\partial F_1}{\partial x} = p_{x_1} \cos \theta - p_{y_1} \sin \theta, 
 \label{F1:momentumx}
\\
p_y &=& \frac{\partial F_1}{\partial y} = p_{x_1} \sin \theta + p_{y_1} \cos \theta. 
\label{F1:momentumy}
\end{eqnarray}
{
Notice that we directly obtain the canonical transformations for $p_{x_1}$ and $p_{y_1}$, 
given by the previous expressions, (\ref{F1:momentumx}) and (\ref{F1:momentumy}), but we 
need to solve equations (\ref{F1:configx}) and (\ref{F1:configy}) to obtain the 
corresponding ones for $x$ and $y$
\begin{eqnarray}
x &=&  {x_1} \cos \theta - y_1 \sin \theta, 
 \label{x:x1y1}
\\
y &=& x_1 \sin \theta + y_1 \cos \theta. 
\label{y:x1y1}
\end{eqnarray}
}

Hence, the first transformed Hamiltonian is 
\begin{multline} \label{ham1a}
H_1= {H}^R + \frac{\partial F_1}{\partial t} = 
\frac{1}{2m} \left(p_{x_1}^2+p_{y_1}^2\right) - \left( \frac{\omega}{2} + 
\dot{\theta}  \right) \left(y_1 p_{x_1} -x_1 p_{y_2} \right) \\
+ \frac{1}{8}\left( m \omega^2+\kappa \right) \left(x_1^2+ y_1^2 \right) 
-qE_x^R x_1  -qE_y^R y_1. 
\end{multline}
Here ${H}^R$ is the original function, but now expressed in terms of the 
transformed coordinates and the rotated electric field 
\begin{eqnarray}
E_x^R &=& E_x \cos \theta + E_y \sin \theta , \label{rot:Ex}\\ 
E_y^R &=& -E_x \sin \theta + E_y \cos \theta . \label{rot:Ey}
\end{eqnarray}
In order to reduce the angular momentum term in Eq. (\ref{ham1a}), we set
\begin{equation}\label{difeq:theta}
\dot{\theta}= -\frac{\omega}{2},
\end{equation} 
and we obtain the direct sum of two one dimensional harmonic oscillator-like
Hamiltonians
\begin{multline} \label{H1}
{H}_1= \frac{1}{2m} \left(p_{x_1}^2+p_{y_1}^2\right) +  \frac{1}{8}\left( m 
\omega^2+\kappa \right) \left(x_1^2+ y_1^2 \right)
-qE_x^R x_1  -qE_y^R y_1.
\end{multline} 
All linear terms can now be 
reduced via space and momentum translations with the generating function 
\begin{equation} \label{F2}
F_2=\left( x_1 -\lambda_x \right)\left( p_{x_2} -\pi_x \right)
+ \left( y_1 -\lambda_y \right)\left(p_{y_2} -\pi_y \right) -S, 
\end{equation} 
that yields the following transformation rules
\begin{eqnarray}
x_2 &=& \frac{\partial F_2}{\partial p_{x_2}}= x_1- \lambda_x, \label{F2:configx}\\
y_2 &=& \frac{\partial F_2}{\partial p_{y_2}}= y_1 -\lambda_y, \label{F2:configy} \\
p_{x_1} &=& \frac{\partial F_2}{\partial x_1}= p_{x_2}-\pi_x, \label{F2:momentumx}\\
p_{y_1} &=& \frac{\partial F_2}{\partial y_1}= p_{y_2}-\pi_y, \label{F2:momentumy}
\end{eqnarray}
where $x_2$, $y_2$, $p_{x_2}$ and $p_{y_2}$ are the new variables and $S$ is the 
action. Here, $\lambda_x$ and $\lambda_y$
 are time dependent parameters for the  translation in coordinates, meanwhile $\pi_x$ 
and  $\pi_y$ are the corresponding ones for the momentum space. { In order 
to obtain the canonical transformations for $x_1$ and $y_1$, we solve 
(\ref{F2:configx}) and (\ref{F2:configy}) 
\begin{eqnarray}
x_1 &=& x_2 +\lambda_x, \label{x1:x2}\\
y_1 &=& y_2 +\lambda_y. \label{y1:y2}
\end{eqnarray}
}
After transforming via $F_2$ the resulting Hamiltonian is 
\begin{multline} \label{ham2l}
H_2 = \frac{1}{2m} \left( p_{x_2}^2 +p_{y_2}^2 \right)
+ \frac{1}{8} \left( m \omega^2 +\kappa \right) \left( x_2^2+y_2^2 \right)\\
- \left( \frac{\pi_x}{m} + \dot{\lambda}_x \right) p_{x_2}
 - \left( \frac{\pi_y}{m} + \dot{\lambda}_y \right) p_{y_2} \\ 
+\left[ \frac{1}{4} \left( m \omega^2 +\kappa \right)\lambda_x -q E_x^R - 
\dot{\pi}_x \right] x_2 +\left[ \frac{1}{4} \left( m \omega^2 +\kappa 
\right)\lambda_y -q E_y^R -\dot{\pi}_y \right] y_2 \\
+\frac{1}{2m} \left( \pi_x^2 +\pi_y^2 \right) +\frac{1}{8} \left( m \omega^2 
+\kappa \right) \left( \lambda_x^2+\lambda_y^2 \right)\\
 -qE_x^R \lambda_x  
-qE_y^R \lambda_y + \dot{\lambda}_x \pi_x + \dot{\lambda}_y \pi_y -\dot{S}. 
\end{multline}

In Hamiltonian $H_2$ the coefficients of $x_2$, $y_2$, $p_{x2}$ and $p_{y2}$ correspond to
the Euler equations of the classical Lagrangian
\begin{multline}
L_1=\frac{1}{2m} \left( \pi_x^2 +\pi_y^2 \right) +\frac{1}{8} \left( m \omega^2 
+\kappa \right) \left( \lambda_x^2+\lambda_y^2 \right)\\
 -qE_x^R \lambda_x  -qE_y^R \lambda_y 
 + \dot{\lambda}_x \pi_x + \dot{\lambda}_y \pi_y 
\end{multline}
for the translation parameters $\lambda_x$,$\lambda_y$, $\pi_x$ and $\pi_y$. 
In order for all the linear coefficients to vanish
we require that this Lagrangian be the solution of the Euler equations for
the translation parameters: 
\begin{eqnarray} 
\frac{d}{dt}\frac{\partial L_1}{\partial\dot \pi_x} - 
\frac{\partial L_1}{\partial \pi_x} &=& 
-\left( \frac{\pi_x}{m} + \dot{\lambda}_x \right) =0 , 
\label{lagrangian:pix}\\
\frac{d}{dt}\frac{\partial L_1}{\partial\dot \pi_y} - 
\frac{\partial L_1}{\partial \pi_y} &=&
-\left( \frac{\pi_y}{m} + \dot{\lambda}_y \right) = 0,
\label{lagrangian:piy} \\
\frac{d}{dt}\frac{\partial L_1}{\partial\dot \lambda_x} - 
\frac{\partial L_1}{\partial \lambda_x} &=&
\frac{ m \omega^2 +\kappa }{4}\lambda_x  -q E_x^R - 
\dot{\pi}_x = 0, 
\label{lagrangian:lambdax}\\
\frac{d}{dt}\frac{\partial L_1}{\partial\dot \lambda_y} - 
\frac{\partial L_1}{\partial \lambda_y} &=&
 \frac{ m \omega^2 +\kappa }{4} \lambda_y -q E_y^R - 
\dot{\pi}_y =0 .
\label{lagrangian:lambday}
\end{eqnarray}
Additionally, to remove the Lagrangian part, $L_1 - \dot S =0$ must be
fulfilled and consequently 
$\dot S$ can be associated with the time derivative of the corresponding
action.

The transformed Hamiltonian $H_2$ is thus simplified into
\begin{equation}
H_2 = \frac{1}{2m} \left( p_{x_2}^2 +p_{y_2}^2 \right) + \frac{1}{8} \left( m 
\omega^2 +\kappa \right) \left( x_2^2+y_2^2 \right). 
\end{equation}
The harmonic oscillator coefficient $m \omega^2 +\kappa$ can be
expressed in terms of new parameters as
\begin{equation}\label{magtime}
m \omega^2 +\kappa = m_0 \omega_0^2{\rm e}^{2\beta -\alpha},
\end{equation}
where ${\rm e}^{2 \beta} = f^2(t) + \kappa_0e^\alpha g(t)/m_0 \omega_0^2$
contains the explicit time dependence of the given magnetic field 
$B = B_0 f(t)$, the confining potential $\kappa = \kappa_0 g(t)$ and 
{
\begin{equation}
m\left(t\right)=m_0{\rm e}^{\alpha\left(t\right)} \label{m_t}.
\end{equation}
}

In terms of this variables, the 
Hamiltonian is rewritten as \cite{comment1}
\begin{equation} \label{H2}
H_2 = \frac{e^{-\alpha}}{2m_0} \left( p_{x_2}^2 +p_{y_2}^2 \right) + \frac{1}{8} 
m_0 \omega_0^2 e^{2 \beta-\alpha} \left( x_2^2+y_2^2 \right). 
\end{equation}

As a next step we consider a dilatation and two shears.
The  generating  function for such a transformation is 
\begin{equation} \label{F3}
F_3 = \frac{e^{\frac{1}{2}\gamma}}{\cos \delta} \left( x_2 p_{x_3}+y_2 p_{y_3} 
\right) - \frac{e^\gamma \tan \delta}{2\Delta} \left( p_{x_3}^2 +p_{y_3}^2 \right) 
- \frac{\Delta \tan \delta}{2} \left( x_2^2 +y_2^2 \right)
\end{equation} 
with time dependent functions $\gamma$, $\delta$ and $\Delta$.
$F_3$ produces the following transformation rules
\begin{eqnarray}
x_3 &=& \frac{\partial F_3}{\partial p_{x_3}} = \frac{e^{\frac{1}{2}\gamma}}{\cos 
\delta} x_2 -\frac{e^\gamma \tan \delta}{\Delta} p_{x_3},
\label{F3:configx} \\ 
y_3 &=& \frac{\partial F_3}{\partial p_{y_3}} = \frac{e^{\frac{1}{2}\gamma}}{\cos 
\delta} y_2 -\frac{e^\gamma \tan \delta}{\Delta} p_{y_3}, 
\label{F3:configy} \\
p_{x_2} &=& \frac{\partial F_3}{\partial x_2} = \frac{e^{\frac{1}{2}\gamma}}{\cos 
\delta} p_{x_3} -\Delta \tan \delta \ x_2, \label{F3:momentumx}\\ 
p_{y_2} &=&  \frac{\partial F_3}{\partial y_2} = \frac{e^{\frac{1}{2}\gamma}}{\cos 
\delta} p_{y_3} -\Delta \tan \delta \ y_2. 
\label{F3:momentumy} 
\end{eqnarray}
{ We can obtain the corresponding canonical transformations by solving 
$x_2$, $y_2$, $p_{x_2}$ and $p_{y_2}$ from the previous equations
\begin{eqnarray}
x_2 &=& e^{-\frac{1}{2}\gamma} x_3 \cos \delta + \frac{e^{\frac{1}{2}\gamma}}{\Delta} p_{x_3} 
\sin \delta \label{x2:x3px3} \\
p_{x_2} &=& e^{\frac{1}{2}\gamma} p_{x_3} \cos \delta -e^{-\frac{1}{2}\gamma} \Delta \ x_3 
\sin \delta \label{px2:x3px3} \\
y_2 &=& e^{-\frac{1}{2}\gamma} y_3 \cos \delta + \frac{e^{\frac{1}{2}\gamma}}{\Delta} p_{y_3} 
\sin \delta \label{y2:y3py3} \\
p_{y_2} &=& e^{\frac{1}{2}\gamma} p_{y_3} \cos \delta -e^{-\frac{1}{2}\gamma} \Delta \ y_3 
\sin \delta \label{py2:y3py3} .
\end{eqnarray}
Notice that even though the generating function $F_3$ in Eq. (\ref{F3}) and 
(\ref{F3:configx})-(\ref{F3:momentumy}) have multiple divergences 
when $\delta = (2n-1)\pi/2$, its corresponding canonical transformation rules
(\ref{x2:x3px3})-(\ref{py2:y3py3}) have non. These are the well known
Arnold transformations\cite{arnold2,aldaya:065302}. It is possible to show that they comply
with condition necessary to preserve the value of the Wronskian
\begin{equation}
\det\left(
\begin{array}{cc}
{\rm e}^{-\frac{1}{2}\gamma} \cos \delta & \frac{e^{\frac{1}{2}\gamma}}{\Delta}\sin \delta\\
-{\rm e}^{-\frac{1}{2}\gamma} \Delta \sin \delta & e^{\frac{1}{2}\gamma} \cos \delta 
\end{array}
\right)=1
\end{equation}
and for the transformation matrix to be symplectic.}

Under $F_3$, the new transformed Hamiltonian is
\begin{multline} \label{H3}
H_3 = \left[ e^{-\alpha} \left( \frac{\Delta}{m_0} \cos^2 \delta +\frac{m_0 
\omega_0^2 e^{2 \beta}}{4 \Delta} \sin^2 \delta \right) -\dot \delta +\sin \delta 
\cos \delta \frac{\dot \Delta}{\Delta}  \right] \frac{e^\gamma}{2 \Delta} 
\left( p_{x_3}^2 +p_{y_3}^2 \right) \\
+ \left[ e^{-\alpha} \left( \frac{\Delta}{m_0} \sin^2 \delta +\frac{m_0 
\omega_0^2 e^{2 \beta}}{4 \Delta} \cos^2 \delta \right) -\dot \delta -\sin \delta 
\cos \delta \frac{\dot \Delta}{\Delta}  \right] \frac{\Delta e^{-\gamma}}{2} 
\left( x_3^2 +y_3^2 \right) \\
+ \left[ e^{-\alpha} \left( -\frac{\Delta}{m_0} +\frac{m_0 \omega_0^2 
e^{2 \beta}}{4 \Delta} \right) \sin \delta \cos \delta +\frac{\dot \gamma}{2}  
-\sin^2 \delta \frac{\dot \Delta}{\Delta}\right] \left( x_3 p_{x_3} +y_3 p_{y_3} 
\right).  
\end{multline}
In order to obtain a null Hamiltonian we set the coefficients
of $\left( p_{x_3}^2 +p_{y_3}^2 \right)$, 
$\left( x_3^2 +y_3^2 \right) $ and  $\left( x_3 p_{x_3} +y_3 p_{y_3} 
\right) $ to zero.  We, thus, obtain the following 
system of coupled differential equations for the transformation parameters 
\begin{eqnarray}
0 & = & e^{-\alpha} \left( \frac{\Delta}{m_0} \cos^2 \delta +\frac{m_0 \omega_0^2 
e^{2 \beta}}{4 \Delta} \sin^2 \delta \right)
-\dot \delta +\sin \delta \cos 
\delta \frac{\dot \Delta}{\Delta} \label{delta1}, \label{kcl1}\\
0 & = & e^{-\alpha} \left( \frac{\Delta}{m_0} \sin^2 \delta +\frac{m_0 \omega_0^2 
e^{2 \beta}}{4 \Delta} \cos^2 \delta \right)
-\dot \delta-\sin \delta \cos 
\delta \frac{\dot \Delta}{\Delta}\label{delta2},\\
0& = & e^{-\alpha} \left( -\frac{\Delta}{m_0} +\frac{m_0 \omega_0^2 e^{2 \beta}}{4 \Delta} 
\right) \sin \delta \cos \delta
+\frac{\dot \gamma}{2}  -\sin^2 \delta 
\frac{\dot \Delta}{\Delta}\label{delta3} .
\end{eqnarray}
The solutions to this differential equations
cancel the whole Hamiltonian $H_3$. In such a case $x_3$, $y_3$, $p_{x_3}$ and $p_{y_3}$ 
are constant in time and, therefore, they are constants of the motion.
To simplify the structure of the differential equations and their solutions we propose
\begin{equation}\label{deltadef}
\Delta = \frac{1}{2} m_0 \omega_0 e^{\beta+\eta},
\end{equation}
where $\eta$ is a time { dependent} function, yielding a simplification of the previous 
coupled equations
\begin{eqnarray}
\dot \delta & = & \frac{1}{2} \omega_0 e^{\beta-\alpha} \cosh \eta , \label{delta4} 
\\
\dot \eta +\dot \beta & = & \frac{1}{2} \omega_0 e^{\beta-\alpha} \sinh \eta 
\left( \tan \delta -\cot \delta \right) ,\label{delta5} \\
\dot \gamma & = & \omega_0 e^{\beta-\alpha} \sinh \eta \tan \delta .\label{delta6}
\end{eqnarray}

For practical purposes the solutions of these equations, in the most general case,
can be obtained by numerical 
methods. Nevertheless, it is possible to extract information from (\ref{delta4})-(\ref{delta6})
by grouping 
the last three equations in a single hyperbolic one
\begin{equation} \label{cuadrature}
\frac{\dot \delta^2}{a^2} -\frac{ \left( \dot \gamma -\dot \beta -\dot \eta  
\right)^2}{b^2} = 1,
\end{equation}
here $a = \omega_0 e^{\beta-\alpha}/2$ and 
$b = \omega_0 e^{\beta-\alpha}/\sin 2 \delta$. If we use the $\eta$ function as 
a parameter, we can rewrite the hyperbola with the parametric functions (\ref{delta4}) and 
\begin{equation}
\dot \gamma  -\dot \beta-\dot \eta = \frac{\omega_0 e^{\beta-\alpha}}{\sin 2 \delta} \sinh \eta .
\end{equation}
For a given problem with no parabolic potential, $\kappa_0 =0$, only one of the branches contains 
the physical solution. Each branch is associated with a given rotating direction
of the charged particle.

In particular, the vertices of the hyperbola correspond to the constant magnetic field case.
If we set ourselves in one of the vertices
$\dot \delta = \omega_0 e^{\beta-\alpha}/2$ and by comparing with (\ref{delta4}) 
we obtain that $\eta = 0$ and, consequently, $\dot \beta = 0$ and $\dot \gamma = 0$. 
This is indeed the case when the magnetic field is a constant, i. e. $\beta = 0$ and 
$\gamma = 0$. In this manner we find that with an appropriate time dependent mass model and 
the initial condition $\delta (0) =0$ we can integrate 
$\dot \delta = \omega_0 e^{-\alpha}/2$ and obtain $\delta$, the only relevant 
parameter under the conditions described above.

It is well-known that the time reversal symmetry is broken by a constant magnetic field, 
even though we have a frictionless problem, this symmetry breaking is the cause of the 
existence two vertices. More generally, for a problem where the magnetic field is a function 
of time, the solution is given by another region at the hyperbola branch. In other words, 
the hyperbolic behavior of Eqs. (\ref{delta4})-(\ref{delta6}) is a consequence of the 
magnetic field's time reversal asymmetry. 

The last transformation gives the solution to 
the initial problem describing the motion of a charged particle 
under the influence of the potentials (\ref{potential:scalar}) and 
(\ref{potential:vector}) where the electric and magnetic fields are only 
time-dependent functions. Under the previous three transformations, we find
that $x_3$, $y_3$, $p_{x_3}$ and $p_{y_3}$ are constants along the classical 
orbit followed by the particle. In other words,  $x_3$, $y_3$, 
$p_{x_3}$ and $p_{y_3}$ are the initial conditions and we shall rename them as $x_3 = 
x_0$, $y_3 = y_0$, $p_{x_3} = p_{x_0}$ and $p_{y_3} = p_{y_0}$. 

It is also possible to figure out a single canonical 
transformation after adequately collecting all the above 
contributions into the following form
\begin{equation} \label{motion}
\xi = {\bf M} \xi_0 +\mu,
\end{equation}
where ${\bf M}$ is a symplectic matrix given by
\begin{multline}\label{simplecticM}
{\bf M} = \left[ \begin{array}{cc}
{\bf a} & {\bf b} \\
{\bf c} & {\bf d}
\end{array} \right] = 
\left[ \begin{array}{cc}
e^{-\frac{\gamma}{2}} \cos \theta \cos \delta & -e^{-\frac{\gamma}{2}} \sin \theta \cos 
\delta   \\
 \frac{e^{\frac{\gamma}{2}}}{\Delta} \sin \theta \sin \delta & 
\frac{e^{\frac{\gamma}{2}}}{\Delta} \cos \theta \sin \delta \\ 
 e^{\frac{\gamma}{2}} \cos \theta \cos \delta & 
-e^{\frac{\gamma}{2}} \sin \theta \cos \delta \\
 e^{\frac{\gamma}{2}} \sin \theta \cos \delta & 
e^{\frac{\gamma}{2}} \cos \theta \cos \delta \\
\end{array} \right.\\
\left. \begin{array}{cc}
\frac{e^{\frac{\gamma}{2}}}{\Delta} \cos \theta \sin \delta & 
-\frac{e^{\frac{\gamma}{2}}}{\Delta} \sin \theta \sin \delta \\
\frac{e^{\frac{\gamma}{2}}}{\Delta} \sin \theta \sin \delta & 
\frac{e^{\frac{\gamma}{2}}}{\Delta} \cos \theta \sin \delta \\ 
e^{\frac{\gamma}{2}} \cos \theta \cos \delta & 
-e^{\frac{\gamma}{2}} \sin \theta \cos \delta \\
e^{\frac{\gamma}{2}} \sin \theta \cos \delta & 
e^{\frac{\gamma}{2}} \cos \theta \cos \delta \\
\end{array} \right]
\end{multline}
and 
\begin{equation}\label{simplectica}
\mu = \left[ \begin{array}{cccc}
\cos \theta & -\sin \theta & 0 & 0 \\
\sin \theta & \cos \theta & 0 & 0 \\
0 & 0 & \cos \theta & -\sin \theta \\
0 & 0 & \sin \theta & \cos \theta \\
\end{array} \right] 
\left[ \begin{array}{c}
\lambda_x \\
\lambda_y \\
-\pi_x \\
-\pi_y 
\end{array} \right].
\end{equation}

As an example,
we consider the simplest case when the magnetic field and the mass are 
constants, meanwhile both the confining potential  and the electric 
field are absent. In such a case $\delta = -\theta =  \omega_0/2 
t$, $\Delta =  m_0 \omega_0/2$ and there is no dilatation, hence 
$\gamma = 0$ and $\eta =0$. 
By using equation (\ref{motion})
all the position and the momentum variables can be expressed as a function
of time and the initial
conditions
\begin{multline}
\left[ \begin{array}{c}
x \\
y \\
p_x \\
p_y 
\end{array} \right] =
\left[ \begin{array}{cc}
\frac{1}{2}\left( 1 + \cos \omega_0 t \right) &
\frac{1}{2}\sin \omega_0 t\\
-\frac{1}{2} \sin \omega_0 t & \frac{1}{2} \left( 1 + \cos \omega_0 t \right) \\ 
-\frac{1}{4} m_0 \omega_0 \sin \omega_0 t & \frac{1}{4} m_0 \omega_0 \left( 
\cos \omega_0 t -1\right)  \\
\frac{1}{4} m_0 \omega_0 \left(1- \cos \omega_0 t \right) & -\frac{1}{4} m_0 
\omega_0 \sin \omega_0 t
\end{array} \right.\\
\left. \begin{array}{cc}
\frac{\sin \omega_0 t}{m_0 \omega_0} & \frac{1-\cos \omega_0 t}{m_0 \omega_0} \\
 \frac{\cos \omega_0 t -1}{m_0 \omega_0} & \frac{\sin \omega_0 t}{m_0 
\omega_0} \\ 
 \frac{1}{2} \left( 1 + \cos \omega_0 t \right) & \frac{1}{2} \sin \omega_0 t \\
 -\frac{1}{2}\sin \omega_0 t &  \frac{1}{2}\left( 1 + \cos \omega_0 t \right) 
\end{array} \right] 
\left[ \begin{array}{c}
x_0 \\
y_0 \\
p_{x_0} \\
p_{y_0} 
\end{array} \right].
\end{multline}
This last result is consistent with the solution obtained directly
from the Hamilton equations of motion.
The motion described in the previous equations is periodic,
with period  $T = 2\pi/\omega_0$ and 
$\omega_0$ is the Larmor frequency. The periodicity can be deduced from the 
behavior of the block matrices ${\bf a}$ and ${\bf d}$
in (\ref{simplecticM}) since they become unit 
matrices for $t = T$, meanwhile ${\bf b}$ and ${\bf c}$ become zero. The charged 
particle is moving around a circular orbit in the plane $xy$ with { radius} 
$r =\sqrt{p_{x_0}^2+p_{y_0}^2}/m_0 \omega_0$. Physically, the 
trajectories of the particles are curved due to the Lorentz force, nevertheless, 
when the magnetic field $B_0$ is small, the motion of the particles is almost linear 
($r$ grows). For larger values of $B_0$, the particle's motion is highly curved 
($r$ decreases). The last feature is given by the off-diagonal block matrices 
${\bf b}$ and ${\bf c}$.

It is important to notice that in the Hamiltonian $H_3$  we can
set the two first coefficients to  $\omega_0/2$ instead of zero as in 
Eqs.(\ref{delta1}) and (\ref{delta2}), meanwhile we keep the null equation
(\ref{delta3}). In this case 
we obtain a KC-like Hamiltonian,
but the equations that must be satisfied in order to obtain a solution
are much more complex.


\section{The quantum problem: unitary transformations}\label{quantization}

The classical calculations presented in the previous section allow
to set a framework for a quantum mechanical 
analog of (\ref{ham2}) through  the Schr\"odinger's equation
\begin{equation}
\hat H\left\vert \psi\left(t\right)\right\rangle=\hat p_t\left\vert \psi\left(t\right)\right\rangle ,
\end{equation}
where, the quantum mechanical Hamiltonian is given by
\begin{equation}\label{qham}
\hat H=\frac{1}{2m}
   \left( \newvec{\hat{p}}-q \newvec{\hat{A}}\right)^2+q\phi
   +\frac{\kappa}{8}\left(\hat x^2+\hat y^2\right).
\end{equation}
Here, $\hat p_t$ is the energy operator, i.e., $\hat p_t\rightarrow i\hbar\partial_t$
and $\hat x$, $\hat y$, $\hat p_x$ and $\hat p_y$ are the
space and momentum operators such that
\begin{eqnarray}
\hat x\left\vert x,y\right\rangle &=& x\left\vert x,y\right\rangle,\\
\hat y\left\vert x,y\right\rangle &=& y\left\vert x,y\right\rangle,\\
\hat p_x\left\vert p_x,p_y\right\rangle &=& p_x\left\vert p_x,p_y\right\rangle,\\
\hat p_y\left\vert p_x,p_y\right\rangle &=& p_x\left\vert p_x,p_y\right\rangle,
\end{eqnarray}
where $\left\vert x,y\right\rangle$ and $\left\vert p_x,p_y\right\rangle$ are the 
space and momentum eigenstates respectively. The space and momentum operators follow the 
usual commutation relations
\begin{equation}
\left[\hat x_i,\hat p_j\right]=i\hbar \delta_{i,j},
\end{equation}
as well as the energy operator and time
\begin{equation}
\left[\hat p_t,t\right]=i\hbar.
\end{equation}
The physical electric and magnetic fields $\newvec{E}$ and $\newvec{B}$, 
respectively, are obtained as usual from the scalar and vector potentials 
$\phi$ and $\newvec{A}$ by the relations (\ref{magneticfield}) and 
(\ref{electricfield}) as we discussed in Sec. \ref{varmas}.

The integration of the quantum mechanical problem follows the same path as
the classical problem. The reduction of the Hamiltonian is now easily achieved
by unitary transformations \cite{messiah,choi1,choi2,choi3}, each one 
associated to one of the three classical canonical transformations applied in 
Sec. \ref{secclas}. Each reduction step has the following structure 
\begin{equation}
U\hat HU^{\dag}U\left\vert\psi\left(t\right)\right\rangle
=U\hat p_tU^{\dag}U\left\vert\psi\left(t\right)\right\rangle
\Rightarrow \hat H'\left\vert\psi'\left(t\right)\right\rangle
=\hat p_t\left\vert\psi'\left(t\right)\right\rangle
\end{equation}
with $\hat H'=U\hat HU^{\dag}-U\left[\hat p_t,U^{\dag}\right]$, and $\left 
\vert\psi'\left(t\right)\right\rangle = U\left\vert\psi\left(t 
\right)\right\rangle$.

The Floquet operator is thus given by
\begin{equation}\label{floquet}
\hat{\mathcal H}=\hat H-\hat p_t,
\end{equation}
and the Schr\"odinger's equation takes the compact form 
\begin{equation}
\hat{\mathcal H}\left\vert\psi\left(t\right)\right\rangle=
\left(\hat H-\hat p_t\right)\left\vert\psi\left(t\right)\right\rangle = 0
\label{shro}.
\end{equation}

Our aim now is to study the Hamiltonian in Eq. (\ref{qham}) for the particular 
case analyzed in Sec. \ref{secclas} of a magnetic and a perpendicular electric  
fields of Eqs. (\ref{magneticfield}) and (\ref{electricfield}). Such fields 
can be obtained from the potentials in (\ref{potential:vector}) and 
(\ref{potential:scalar}). The quantum mechanical potentials are thus given by
\begin{eqnarray}
\newvec{A}&=&\frac{B}{2}\left(- \hat y\ui+\hat x\uj\right), \\
\phi &=& -E_x\left(t\right)\hat x
         -E_y\left(t\right)\hat y.
\end{eqnarray}
In this gauge, the Floquet operator takes the following form
\begin{multline}
\hat{ \mathcal{H}}=\frac{1}{2m}
\left[
   \left(\hat p_x+\frac{qB}{2}\hat y\right)^2
  +\left(\hat p_y-\frac{qB}{2}\hat x\right)^2
\right]\\
  -q\left[E_x\left(t\right)\hat x+E_y\left(t\right)\hat y\right]
  +\frac{\kappa}{8}\left(\hat x^2+\hat y^2\right)
  -\hat p_t.
\end{multline}


\subsection{Evolution operator}\label{unitoperator}

To obtain the evolution operator for the Hamiltonian in (\ref{qham}) in the 
presence of the magnetic and electric fields given by Eqs. 
(\ref{magneticfield}) and (\ref{electricfield}), respectively, we proceed in a 
similar fashion to the classical case in Sec. \ref{secclas}. We apply a series 
of unitary transformations, each corresponding to a canonical transformation of 
the classical case.

The first unitary transformation, a rotation around the $z$ axis \cite{messiah},
corresponds to the canonical transformation in Eq. (\ref{F1}) and is given by
\begin{equation}
U_{1}=\exp\left(i\frac{\theta \hat L_z}{\hbar} \right),
\end{equation}
where $\hat L_z=\hat x \hat p_y-\hat y \hat p_x$ is the angular momentum
along the $z$ axis.
It has the following effect on the position, momentum and energy operators
\begin{eqnarray}
U_{1}\hat x U_{1}^{\dag} &=& \hat x \cos\theta-\hat y\sin\theta,\label{trans1:a}\\
U_{1}\hat y U_{1}^{\dag} &=& \hat x \sin\theta+\hat y\cos\theta,\\
U_{1}\hat p_x U_{1}^{\dag} &=& \hat p_x \cos\theta-\hat p_y\sin\theta,\\
U_{1}\hat p_y U_{1}^{\dag} &=& \hat p_x \sin\theta+\hat p_y\cos\theta,\\
U_{1}\hat p_t U_{1}^{\dag} &=& \hat p_t+\dot \theta \hat L_z.\label{trans1:b}
\end{eqnarray}
Note that $U_1$ leaves invariant the quadratic forms $\hat x^2+\hat y^2$ and 
$\hat p_x^2+\hat p_y^2$, yielding the transformed Floquet operator
\begin{multline}
U_{1}HU_{1}^{\dag} = \frac{1}{2m}
  \left(\hat p_x^2+\hat p_y^2\right)
  +\frac{1}{8}\left(m\omega^2+\kappa\right)
  \left(\hat x^2+\hat y^2\right)\\
  -q\left(E_x^R\hat x
         +E_y^R\hat y\right)
         -\left(\dot \theta+\frac{qB}{2m}\right)\hat L_z-\hat p_t,
\end{multline}
where $E_x^R$ and $E_y^R$ are the rotated components of the electric field 
given in Eqs. (\ref{rot:Ex}) and (\ref{rot:Ey}). If the time dependent 
parameter  $\theta$ of the $U_{1}$ transformation follows Eq. 
(\ref{difeq:theta}) it is possible to reduce the term proportional to the 
angular momentum $\hat L_z$. The Floquet operator is thus completely separated 
into the $x$ and $y$ parts. Now the Schr\"odinger equation takes the shape of two 
uncoupled one dimensional harmonic oscillators. Here it is important to set 
$\theta\left(0\right)=0$ as the initial condition for the parameter in order that
$U_{1}$ goes to unity as $t\rightarrow 0$. We will set this initial condition 
for all the transformations' parameters. Once the Floquet operator is separated 
we can proceed to reduce each part with the unitary transformations. We note 
that if $m$ is time independent then $\theta = -\omega_0t/2$ with $\omega_0=qB_0/m_0$, 
the cyclotron frequency. In this case, the charged particle motion is taken to a 
reference system that turns at half the cyclotron angular frequency.

The next unitary transformation corresponds to displacements in space, 
momentum and energy and is associated with the canonical transformation
(\ref{F2}). It is given by
\begin{eqnarray}
U_{2}=U_{2t}U_{2x}U_{2y},\label{wtransform}
\end{eqnarray}
where
\begin{eqnarray}
U_{2t} &=& \exp\left[\frac{i}{\hbar}S\left(t\right)\right],\\
U_{2x} &=& \exp\left[\frac{i}{\hbar}\pi_x\left(t\right)\hat x\right]
        \exp\left[\frac{i}{\hbar} \lambda_x\left(t\right)\hat p_x\right],\\
U_{2y} &=& \exp\left[\frac{i}{\hbar}\pi_y\left(t\right)\hat y\right]
        \exp\left[\frac{i}{\hbar} \lambda_y\left(t\right)\hat p_y\right],
\end{eqnarray}
with time dependent transformation parameters
 $S\left(t\right)$, $ \lambda_x\left(t\right)$, $\pi_x\left(t\right)$,
$ \lambda_y\left(t\right)$ and $\pi_y\left(t\right)$.
This unitary operator
(\ref{wtransform}) yields the following
transformation rules
\begin{eqnarray}
U_{2}\hat xU_{2}^{\dag}   &=& \hat x    +  \lambda_x\left(t\right)\label{trans2:a},\\
U_{2}\hat p_xU_{2}^{\dag} &=& \hat p_x  - \pi_x\left(t\right), \\
U_{2}\hat yU_{2}^{\dag}   &=& \hat y    +  \lambda_y\left(t\right), \\
U_{2}\hat p_yU_{2}^{\dag} &=& \hat p_y  - \pi_y\left(t\right),\\
U_{2}\hat p_tU_{2}^{\dag} &=& \hat p_t +\dot S-\dot \lambda_x\pi_x-\dot \lambda_y\pi_y\nonumber\\
              &&+\dot \pi_x \hat x+\dot \lambda_x\hat p_x
                +\dot \pi_y \hat y+\dot \lambda_y\hat p_y.\label{trans2:b}
\end{eqnarray}
Now we apply successively transformations $U_1$ and $U_2$ to the Floquet operator
(\ref{floquet}) obtaining
\begin{multline}
U_{2}U_{1}\mathcal{H}U_{1}^{\dag}U_{2}^{\dag}
=\frac{1}{2m}
  \left[\left(\hat p_x-\pi_x\right)^2+\left(\hat p_y-\pi_y\right)^2
  \right]\\
  +\frac{1}{8}\left(m\omega^2+\kappa\right)
      \left(\hat x+ \lambda_x\right)^2
 +\frac{1}{8}\left(m\omega^2+\kappa\right)
      \left(\hat y+\lambda_y\right)^2 \\
  -q\left[E_x^R\left(\hat x+ \lambda_x\right)
         +E_y^R\left(\hat y+ \lambda_y \right)
    \right]-\hat p_t -\dot S\left(t\right)\\
        +\dot \lambda_x\pi_x+\dot \lambda_y\pi_y
         -\dot \pi_x \hat x-\dot \pi_y\hat y-\dot \lambda_x\hat p_x-\dot \lambda_y\hat p_y.
         \label{transu2u1}
\end{multline}
In the previous transformed Floquet operator, as in the classical Hamiltonian,
we identify the Lagrangian $L_1$ 
of the transformation parameters and the
corresponding Euler equations (\ref{lagrangian:pix})-(\ref{lagrangian:lambday}). 
Eq. (\ref{transu2u1}) can be recast in the following form
\begin{multline}
U_{1}U_{2}\mathcal{H}U_{1}^{\dag}U_{2}^{\dag}
= \frac{1}{2m}
  \left(\hat p_x^2+\hat p_y^2\right)
  +\frac{1}{8}\left(m\omega^2+\kappa\right)
  \left(\hat x^2+\hat y^2\right)\\
 - \left[\frac{d}{dt}\frac{\partial L_1}{\partial\dot \lambda_x}
   - \frac{\partial L_1}{\partial \lambda_x}\right]\hat x
   + \left[\frac{d}{dt}\frac{\partial L_1}{\partial\dot \pi_x}
   - \frac{\partial L_1}{\partial \pi_x}\right]\hat p_x\\
   - \left[\frac{d}{dt}\frac{\partial L_1}{\partial\dot \lambda_y}
   - \frac{\partial L_1}{\partial \lambda_y}\right]\hat y
   + \left[\frac{d}{dt}\frac{\partial L_1}{\partial\dot \pi_y}
   - \frac{\partial L_1}{\partial \pi_y}\right]\hat p_y\\
    -p_t+L-\dot S.
\end{multline}
In order to reduce the linear terms and simplify the Floquet operator,
we assume that
the Euler equations (\ref{lagrangian:pix})-(\ref{lagrangian:lambday})
are met for the parameters
$ \lambda_x$, $ \lambda_y$, $\pi_x$, $\pi_y$ and $S$.
The transformed Floquet operator (\ref{transu2u1}) is thus simplified into
\begin{equation}
U_{1}U_{2}\mathcal{H}U_{1}^{\dag}U_{2}^{\dag} = \frac{1}{2m}
  \left(\hat p_x^2+\hat p_y^2\right)\\
  +\frac{1}{8}\left(m\omega^2+\kappa\right)
  \left(\hat x^2+\hat y^2\right)
  -\hat p_t.
\end{equation}

Corresponding to the $F_3$ canonical transformation in Eq. (\ref{F3}),
the last unitary transformation can be split into the $x$ and $y$ parts
as shown below
\begin{equation}
U_3=U_{3x}U_{3y}.\label{u3}
\end{equation}
The first unitary transformations in the right hand side
is devoted to reducing
the quadratic terms in the $x$ part of the Floquet operator
and correspondingly the second term reduce the $y$ part of
the Hamiltonian. 

The first transformation corresponds to a shear
and is given by
{
\begin{equation}
U_{3x}
=\exp\left[-i\frac{\gamma}{4\hbar}
\left(\hat x\hat p_x+\hat p_x\hat x\right)\right]
\exp\left[i\frac{\delta}{2\hbar}
\left(\Delta \hat x^2+\frac{1}{\Delta}\hat p_x^2\right)\right]
,\label{u3a}
\end{equation}
}
and yields the following transformation rules
\begin{eqnarray}
U_{3x}\hat xU_{3x}^\dag
&=& {\rm e}^{-\frac{\gamma}{2}}\hat x \cos\delta
+\frac{{\rm e}^{\frac{\gamma}{2}}}{\Delta}\hat p_x\sin\delta,\label{trans3:a}\\
U_{3x}\hat p_xU_{3x}^\dag
&=& {\rm e}^{\frac{\gamma}{2}}\hat p_x\cos\delta
-{\rm e}^{-\frac{\gamma}{2}}\Delta\hat x\sin\delta,\\
U_{3x}\hat p_tU_{3x}^\dag
&=& \hat p_t
  +\left(\frac{\dot\Delta}{2\Delta}\sin^2\delta-\frac{\dot\gamma}{4}\right)
  \left(\hat x\hat p_x+\hat p_x\hat x\right)\nonumber \\
    &&+\frac{{\rm e}^{\gamma}}{2\Delta}\left(\dot\delta
    -\frac{\dot\Delta}{2\Delta}\sin 2\delta\right)\hat p_x^2\nonumber\\
&&+\frac{\Delta{\rm e}^{-\gamma}}{2}\left(\dot\delta
  +\frac{\dot\Delta}{2\Delta}\sin 2\delta\right)\hat x^2\label{trans3:b}.
\end{eqnarray}
{
The first two equations are in fact the quantum version of the
Arnold transformation\cite{arnold2,aldaya:065302}
as pointed out in Sec. \ref{secclas}.
In order to compute Eq. (\ref{trans3:b}) it is necessary to obtain the
time derivative of the unitary transformation. One way to perform this
derivative would be to use the Magnus formula \cite{pechukas:3897,blanes:151}
since it can not be computed by
direct derivation because the generator
of (\ref{u3a}) does not necessarily commute with its time derivative.
Nevertheless we follow an alternative method by
separating the transformation (\ref{u3a}) into
a shear and a dilatation
\begin{multline}
U_{3x}=
\exp\left[-\frac{i\gamma}{4\hbar}\left(\hat x\hat p_x+\hat p_x\hat x\right)\right]
\exp\left[\frac{i\mu}{2\hbar}\left(\hat x\hat p_x+\hat p_x\hat x\right)\right]\\
\times
\exp\left[\frac{i \delta}{2\hbar}\left(\Delta_0 \hat x^2+\frac{1}{\Delta_0}\hat p_x^2\right)\right]
\exp\left[-\frac{i\mu}{2\hbar}\left(\hat x\hat p_x+\hat p_x\hat x\right)\right]
,
\end{multline}
}
where $\Delta=\Delta_0e^{2\mu}$ and $\Delta_0$ is a constant.
Here it is convenient to set the time dependence of the mass, magnetic
field and confining potential by means of (\ref{magtime}) and 
{ (\ref{m_t}).} 
Applying this transformation to the Floquet operator we readily obtain
\begin{multline}
U_{3x}U_{2}U_{1}
\mathcal{H}U_{1}^{\dag}U_{2}^{\dag}U_{3x}^\dag\\
   =\left[{\rm e}^{-\alpha}\left(\frac{\Delta}{m_0}\cos^2\delta
  +\frac{m_0\omega_0^2{\rm e}^{2\beta}}{4\Delta}\sin^2\delta\right)
  -\dot\delta+\frac{\dot \Delta}{\Delta}\sin\delta\cos\delta \right]
  \frac{{\rm e}^\gamma}{2\Delta}\hat p_x^2\\
  +\left[{\rm e}^{-\alpha}\left(\frac{\Delta}{m_0}\sin^2\delta
  +\frac{m_0\omega_0^2{\rm e}^{2\beta}}{4\Delta}\cos^2\delta\right)
  -\dot\delta-\frac{\dot \Delta}{\Delta}\sin\delta\cos\delta\right]
  \frac{\Delta{\rm e}^{-\gamma}}{2}\hat x^2\\
   +\left[{\rm e}^{-\alpha}\left(-\frac{\Delta}{m}
  +\frac{m\omega_0^2{\rm e}^{2\beta}}{4\Delta}\right)\sin\delta\cos\delta
  -\frac{\dot \Delta}{\Delta}\sin^2\delta+\frac{\dot \gamma}{2}\right]
  \frac{1}{2}\left(\hat x \hat p_x+\hat p_x \hat x\right)\\
   +\frac{1}{2m}\hat p_y^2
  +\frac{1}{8}\left(m\omega^2+\kappa\right)\hat y^2-\hat p_t.
\end{multline}
In order to vanish the terms proportional to $\hat x\hat p_x+\hat p_x \hat x$,
$\hat p_x^2$ and $\hat x^2$, the differential equations for the $\gamma$, $\Delta$ and $\delta$
parameters between parenthesis
should vanish. Notice that this equations are the same as Eqs. (\ref{delta1})-(\ref{delta2})
and consequently to Eqs. (\ref{delta4})-(\ref{delta6}). Lastly the Floquet operator
reduces to
\begin{equation}
U_{3x}U_{2}U_{1}
\mathcal{H}U_{1}^{\dag}U_{2}^{\dag}U_{3x}^\dag
  =\frac{1}{2m}\hat p_y^2
  +\frac{1}{8}\left(m\omega^2+\kappa\right)\hat y^2-\hat p_t.
\end{equation}

The $y$ part of the Hamiltonian can be eliminated by
a transformation similar to Eq. (\ref{u3a}) given by
{
\begin{equation}
U_{3y}
=\exp\left[-i\frac{\gamma}{4\hbar}
\left(\hat y\hat p_y+\hat p_y\hat y\right)\right]
\exp\left[i\frac{\delta}{2\hbar}
\left(\Delta \hat y^2+\frac{1}{\Delta}\hat p_y^2\right)\right].
\end{equation}
}
In this transformation the parameters $\gamma$, $\Delta$ and $\delta$
are the same as those from  Eq. (\ref{u3a}) since the
$x$ and $y$ parts of the Hamiltonian are symmetrical.
By applying this transformation we finally obtain
\begin{equation}\label{qevop}
\tilde{\mathcal{H}}=U_{3}U_{2}U_{1}
\mathcal{H}U_{1}^{\dag}U_{2}^{\dag}U_{3}^\dag
=-\hat p_t.
\end{equation}
In Eq. (\ref{qevop}), the Floquet operator was reduced to the energy operator $\hat p_t$
implying that any \emph{ket} applied to the right of $\tilde{\mathcal{H}}$ as
\begin{equation}
\tilde{\mathcal{H}}U_{3}U_{2}U_{1}
\left\vert\psi\left(t\right)\right\rangle=-\hat p_tU_{3}U_{2}U_{1}\left\vert\psi\left(t\right)\right\rangle=0,
\end{equation}
 should be 
a constant one, according with Schr\"odinger's equation (\ref{shro}), e.g.,
\begin{equation}
U_{3}U_{2}U_{1}\left\vert\psi\left(t\right)\right\rangle=\left\vert\psi\left(0\right)\right\rangle.
\end{equation}
As a consequence, the state of the system at any time $\left\vert\psi\left(t\right)\right\rangle$
is connected to the state in $t=0$, $\left\vert\psi\left(0\right)\right\rangle$, by
\begin{equation}
\left\vert\psi\left(t\right)\right\rangle=U_{1}^{\dag}U_{2}^{\dag}U_{3}^\dag
\left\vert\psi\left(0\right)\right\rangle .
\end{equation}
The time evolution operator is thus easily obtained as
\begin{equation}\label{utrans}
\mathcal{U}\left(t_1,t_0\right)=U_{1}^{\dag}\left(t_1,t_0\right)
U_{2}^{\dag}\left(t_1,t_0\right)U_{3}^\dag\left(t_1,t_0\right) .
\end{equation}
and the state of the system at any given time $t_1$ evolves
from $t_0$ according with
\begin{equation}
\left\vert\psi\left(t_1\right)\right\rangle
=\mathcal U\left(t_1,t_0\right)\left\vert\psi\left(t_0\right)\right\rangle .
\end{equation}
Notice that time enters the evolution operator $\mathcal U$ through the
parameters $\theta$, $ \lambda_x$, $\pi_x$, $ \lambda_y$, $\pi_y$,
$\gamma$, $\delta$ and $\Delta$
in each of the unitary transformations.

{ It is easy to verify that the obtained unitary transformation $\mathcal U$
corresponds to the Magnus expansion\cite{pechukas:3897,blanes:151} of
the Dyson series at first order
\begin{equation}
\mathcal{U}\equiv 1-\frac{i}{\hbar}\int_{t_0}^{t_1} dtH\left(t\right).
\end{equation}
}

We can calculate the position and momentum operators in the Heisenberg
representation by performing the three transformations on
Schr\"odinger representation of the space an momentum operators
\begin{eqnarray}
\hat x_H\left(t\right) &=&\mathcal{U}^{\dag}\hat x\mathcal{U}=U_3U_2U_1\hat x U_{1}^{\dag}U_{2}^{\dag}U_{3}^\dag,\\
\hat y_H\left(t\right) &=&\mathcal{U}^{\dag}\hat y\mathcal{U}=U_3U_2U_1\hat y U_{1}^{\dag}U_{2}^{\dag}U_{3}^\dag,\\
\hat p_{xH}\left(t\right) &=&\mathcal{U}^{\dag}\hat p_x\mathcal{U}=U_3U_2U_1\hat p_x U_{1}^{\dag}U_{2}^{\dag}U_{3}^\dag,\\
\hat p_{yH}\left(t\right) &=&\mathcal{U}^{\dag}\hat p_y\mathcal{U}=U_3U_2U_1\hat p_y U_{1}^{\dag}U_{2}^{\dag}U_{3}^\dag.
\end{eqnarray}
By working out the explicit form of the previous transformations and using
the transformation rules (\ref{trans1:a})-(\ref{trans1:b}), (\ref{trans2:a})-(\ref{trans2:b})
and (\ref{trans3:a})-(\ref{trans3:b}) we obtain
{
\begin{eqnarray}
\hat x_H\left(t\right) &=&
    {\rm e}^{-\frac{\gamma}{2}}\cos\theta\cos\delta \hat x
   -{\rm e}^{-\frac{\gamma}{2}}\sin\theta\cos\delta \hat y\nonumber \\
   &&+\frac{{\rm e}^{\frac{\gamma}{2}}}{\Delta}\cos\theta\sin\delta \hat p_x
   -\frac{{\rm e}^{\frac{\gamma}{2}}}{\Delta}\sin\theta\sin\delta \hat p_y
   +\lambda_x \cos\theta-\lambda_y\sin\theta,\\
\hat y_H\left(t\right) &=&
    {\rm e}^{-\frac{\gamma}{2}}\sin\theta\cos\delta \hat x
   +{\rm e}^{-\frac{\gamma}{2}}\cos\theta\cos\delta \hat y\nonumber \\
 &&+\frac{{\rm e}^{\frac{\gamma}{2}}}{\Delta}\sin\theta\sin\delta \hat p_x
   +\frac{{\rm e}^{\frac{\gamma}{2}}}{\Delta}\cos\theta\sin\delta \hat p_y
   +\lambda_x\sin\theta+\lambda_y\cos\theta,\\
\hat p_{xH}\left(t\right) &=&
   -{\rm e}^{-\frac{\gamma}{2}}\Delta\cos\theta\sin\delta \hat x
   +{\rm e}^{-\frac{\gamma}{2}}\Delta\sin\theta\sin\delta \hat y\nonumber \\
 &&+{\rm e}^{\frac{\gamma}{2}}\cos\theta\cos\delta \hat p_x
   -{\rm e}^{\frac{\gamma}{2}}\sin\theta\cos\delta \hat p_y
   -\pi_x\cos\theta+\pi_y\sin\theta,\\
\hat p_{yH}\left(t\right) &=&
   -{\rm e}^{-\frac{\gamma}{2}}\Delta\sin\theta\sin\delta \hat x
   -{\rm e}^{-\frac{\gamma}{2}}\Delta\cos\theta\sin\delta \hat y\nonumber \\
 &&+{\rm e}^{\frac{\gamma}{2}}\sin\theta\cos\delta \hat p_x
   +{\rm e}^{\frac{\gamma}{2}}\cos\theta\cos\delta \hat p_y
   -\pi_x\sin\theta-\pi_y\cos\theta.
\end{eqnarray}
}
Here, it is worthwhile noticing that, as in the classical case,
the previous equations can be
also expressed in the symplectic form as
\begin{equation}\label{qsimplectic}
\left[\begin{array}{c}
\hat x_H\\
\hat y_H\\
\hat p_{xH}\\
\hat p_{yH}\\
\end{array}\right]=\mathbf{M}\left[\begin{array}{c}
\hat x\\
\hat y\\
\hat p_x\\
\hat p_y\\
\end{array}\right]+\mu,
\end{equation}
where $\mathbf{M}$ and $\mu$ are given by Eqs. (\ref{simplecticM})
and (\ref{simplectica}), respectively.

\subsection{Green's function}\label{green}

The Green's function is calculated as usual in terms of the evolution
operator as
\begin{equation}
G\left(x,y,t\left\vert\right.x^\prime,y^\prime,0\right)
  =\left\langle x,y\left\vert\mathcal{U}\left(t\right)\right\vert x^\prime,y^\prime\right\rangle.
\end{equation}
To obtain the explicit form of $G$ we first calculate the matrix
elements of each of the unitary transformations $U_1$, $U_2$ and $U_3$
in order to join them by the integral
\begin{multline}
G\left(x,y,t\left\vert\right.x^\prime,y^\prime,0\right)
=\int\int dx_1^2dx_2^2\left\langle x,y\left\vert U_1^\dag\right\vert x_1,y_1\right\rangle\\
\times\left\langle x_1,y_1\left\vert U_2^\dag\right\vert x_2,y_2\right\rangle
\left\langle x_2,y_2\left\vert U_3^\dag\right\vert x^\prime,y^\prime\right\rangle .
\label{integralg}
\end{multline}

For $U_1$ and $U_2$ it is convenient to explore their
effect on a space eigenstate.
The rotation $U_1$ has the expected effect on any space \emph{eigenket}
\begin{equation}
U_1^\dag\left\vert x,y \right\rangle
=\left\vert x\cos\theta-y\sin\theta,y\cos\theta+x\sin\theta \right\rangle, 
\end{equation}
and its matrix element is hence given by
\begin{equation}
\left\langle x^{\prime},y^{\prime}\left\vert U_1^\dag\right\vert x,y \right\rangle
= \delta\left(x^{\prime}-x\cos\theta+y\sin\theta\right)\\
   \delta\left(y^{\prime}-y\cos\theta-x\sin\theta\right).
\end{equation}
The transformation $U_2$ is a translation in space and momentum,
therefore its effect on a
space \emph{eigenstate} is
\begin{equation}
U_2^\dag\left\vert x,y \right\rangle={\rm e}^{-i\frac{\pi_x x}{\hbar}}{\rm e}^{-i\frac{\pi_y y}{\hbar}}
\left\vert x +\lambda_x,y +\lambda_y\right\rangle,
\end{equation}
and its matrix element is thus given by
\begin{equation}
\left\langle x^{\prime},y^{\prime} \left\vert U_2^{\dag}
  \right\vert x,y\right\rangle=
  {\rm e}^{-i\frac{\pi_xx}{\hbar}}{\rm e}^{-i\frac{\pi_y y}{\hbar}}
  \times\delta\left( x^{\prime}-x-\lambda_x\right) 
  \delta\left(y^{\prime}-y-\lambda_y\right).
\end{equation}

The transformation $U_3$ is the product of a dilatation and a shear.
The dilatation has the following effect on a space \emph{eigenstate}
\begin{equation}
{\rm e}^{-i\frac{\gamma}{4\hbar}\left(\hat x\hat p_x+\hat p_x \hat x\right)}\left\vert x\right\rangle
={\rm e}^{-\frac{\gamma}{4}}\left\vert x{\rm e}^{-\frac{\gamma}{2}}\right\rangle,
\end{equation}
where the coefficient ${\rm e}^{-\frac{\gamma}{4}}$ is due to the rescaling
of space and the consequent renormalization of the space \emph{eigenket}.
The matrix element of the dilatation is then given by
\begin{equation}
\left\langle x^{\prime}\left\vert
{\rm e}^{-i\frac{\gamma}{4\hbar}\left(\hat x\hat p_x+\hat p_x \hat x\right)}
\right\vert x\right\rangle
={\rm e}^{-\frac{\gamma}{4}}\delta\left(x^{\prime}- x{\rm e}^{-\frac{\gamma}{2}}\right).
\end{equation}
The shear matrix element is in fact the propagator for an harmonic
oscillator \cite{sakurai}
\begin{equation}
\left\langle x^{\prime}\left\vert
{\rm e}^{i\frac{\delta}{2\hbar}\left(\Delta\hat x^2+\frac{1}{\Delta}\hat p_x^2 \right)}
\right\vert x\right\rangle
=\sqrt{\frac{\Delta}{2\pi\hbar \sin\delta}}
{\rm e}^{-\frac{i\Delta}{2\hbar\sin\delta}\left[2xx^{\prime}
 -\left(x^2+x^{\prime 2}\right)\cos\delta\right]}.
\end{equation}

After reducing all the integrals in (\ref{integralg})
the explicit form for the Green's function is obtained as
\begin{multline}
G\left(x,y,t\left\vert\right.x^\prime,y^\prime,0\right)=
\frac{\Delta {\rm e}^{-\frac{\gamma}{2}}}{2\pi\hbar \sin\delta}
  {\rm e}^{-\frac{i}{\hbar}S\left(t\right)} 
  {\rm e}^{\frac{i\Delta {\rm e}^{-\gamma}}{2\hbar}\cot\delta
      \left(x^{\prime 2}+y^{\prime 2}\right)}\\
\times{\rm e}^{\frac{i\Delta}{2\hbar}\cot\delta\left[
  \left(x\cos\theta+y\sin\theta-\lambda_x\right)^2+
  \left(y\cos\theta-x\sin\theta-\lambda_y\right)^2\right]}\\
\times {\rm e}^{-\frac{i}{\hbar}
  \left(\pi_x + \frac{\Delta}{\sin\delta}x'{\rm e}^{-\gamma/2}\right)
  \left(x\cos\theta+y\sin\theta-\lambda_x\right)}\\
\times{\rm e}^{-\frac{i}{\hbar}\left(\pi_y+\frac{\Delta}{\sin\delta}y'
      {\rm e}^{-\gamma/2}\right)\left(y\cos\theta-x\sin\theta-\lambda_y\right)}.
\end{multline}
This Green's function has indeed the correct shape predicted
by Schwinger and others\cite{schw1,schw2,schw3};
it should be composed only of linear and quadratic
terms of the space and momentum operators.

\subsection{Gaussian Wave Packet}\label{gaussian}

As an example, we now wish to study the evolution of a charged particle
Gaussian wave packet under the action
of constant and uniform crossed electric and magnetic fields.
For the mass we select the LTDMM from  Eq. (\ref{masdep}), and we
set the same parameters from Sec. \ref{varmas} in order to
prove Ehrenfest theorem.

We start (in $t=0$) with a Gaussian wave packet of the form
\begin{equation}
\psi\left(x,y,0\right)=\frac{1}{\sqrt{\pi}a}
\exp\left(-\frac{x^2+y^2}{2 a^2}\right)
\exp\left(i\frac{p_{x0}x+p_{y0}y}{\hbar}\right),
\end{equation}
where $p_{x0}$, $p_{y0}$ and $a$ are the initial momentum and wave packet width values.
Note that initially the wave packet's center is located at the origin, and since the
constant magnetic field is directed along the $z$ axis, the vector potential is given
by (\ref{potential:vector}) therefore its average vanishes. In this manner, the initial momentum
and velocity are related by
$p_{x0}\ui + p_{y0}\uj=m \boldsymbol{\dot{r}}\left(0\right)$.

The wave function at any time is explicitly calculated
in terms of the transformation parameters as
\begin{multline}
\psi\left(x,y,t\right)=\int dx^{\prime}dy^{\prime}
G\left(x,y,t\left\vert\right.x^\prime,y^\prime,0\right)\psi\left(x^\prime,y^\prime,0\right)\\
=\frac{{\rm e}^{-\frac{i}{\hbar}S\left(t\right)}}{\sqrt{\pi}\sigma^2}\left(\frac{\hbar{\rm e}^{\frac{\gamma}{2}}\sin\delta}{a\Delta}
+ia{\rm e}^{-\frac{\gamma}{2}}\cos\delta\right)\\
\times\exp\left\{-\frac{1}{2\sigma^2}\left[
\left(x-\zeta_x^R\right)^2+
\left(y-\zeta_y^R\right)^2\right]\right\}\\
\times\exp\left\{-i\frac{a^2{\rm e}^{-\gamma}}{2\pi\sigma^2}
\cot\delta\left[
\left(x-\zeta_x^R\right)^2+
\left(y-\zeta_y^R\right)^2\right]\right\}\\
\times\exp\left\{i\frac{\Delta\cot\delta}{2\hbar}
\left[
\left(x-\lambda_x^R\right)^2+\left(y-\lambda_y^R\right)^2
\right]\right\}\\
\times\exp\left\{
-\frac{i}{\hbar}\left[
\pi_x^R\left(x-\lambda_x^R\right)+\pi_y^R\left(y-\lambda_y^R\right)
\right]
\right\},
\label{packet}
\end{multline}
where the standard deviation of the wave packet is
given by
\begin{equation}
\sigma\left(t\right)=\sqrt{a^2{\rm e}^{-\gamma}\cos^2\delta+
  \frac{\hbar^2{\rm e}^\gamma\sin^2\delta}{a^2\Delta^2}},
\end{equation}
and correctly complies with $\sigma\left(0\right)=a$.
The rotated $\lambda$, $\pi$ and $\zeta$ parameters are given by
\begin{eqnarray}
\lambda_x^R &=& \lambda_x\cos\theta-\lambda_y\sin\theta,\\
\lambda_y^R &=& \lambda_y\cos\theta+\lambda_x\sin\theta,\\
\pi_x^R &=& \pi_x\cos\theta-\pi_y\sin\theta,\\
\pi_y^R &=& \pi_y\cos\theta+\pi_y\sin\theta,\\
\zeta_x^R &=& \zeta_x\cos\theta-\zeta_y\sin\theta,\\
\zeta_y^R &=& \zeta_y\cos\theta+\zeta_x\sin\theta.
\end{eqnarray}
The $\zeta$ parameters are composed of two parts
\begin{eqnarray}
\zeta_x &=& \lambda_x+\lambda_{x0},\\
\zeta_y &=& \lambda_y+\lambda_{y0},
\end{eqnarray}
where
\begin{eqnarray}
\lambda_{x0} &=& \frac{{\rm e}^{\frac{\gamma}{2}}}{\Delta\csc\delta}p_{x0},\\
\lambda_{y0} &=& \frac{{\rm e}^{\frac{\gamma}{2}}}{\Delta\csc\delta}p_{y0}.
\end{eqnarray}
The probability density can easily be worked out from Eq. (\ref{packet})
giving
\begin{equation}
\left\vert \psi\left(x,y,t\right) \right\vert^2
=\frac{1}{\pi \sigma^2} \exp\left[
-\frac{1}{\sigma^2}\left(x-\zeta_x^R\right)^2\right]
\exp\left[
-\frac{1}{\sigma^2}\left(y-\zeta_y^R\right)^2\right].\label{probdens}
\end{equation}

From the previous expression it is clear that
the Gaussian wave packet follows
the trajectory given by the vector $\boldsymbol{\zeta}=\zeta_x^R\ui + \zeta_y^R\uj$.
Moreover, using  the differential Eqs.(\ref{difeq:theta}), (\ref{deltadef})-(\ref{delta6}) and
(\ref{lagrangian:pix})-(\ref{lagrangian:lambday}) it is easily demonstrated
that  $\lambda_x^R$ and $\lambda_y^R$ fulfill the same equations
of motion as the classical particle
\begin{eqnarray}
m \ddot\lambda_x^R-m\omega\dot \lambda_y^R+\dot m\dot \lambda_x^R-q E_x=0,\label{lambdaxR}\\
m \ddot\lambda_y^R+m\omega\dot \lambda_x^R+\dot m\dot \lambda_y^R-q E_y=0,
\end{eqnarray}
and $\lambda_{x0}^R$ and $\lambda_{y0}^R$ fulfill the homogeneous equations
\begin{eqnarray}
m \ddot\lambda_{x0}^R-m\omega\dot \lambda_{y0}^R+\dot m\dot \lambda_{x0}^R=0,\\
m \ddot\lambda_{y0}^R+m\omega\dot \lambda_{x0}^R+\dot m\dot \lambda_{y0}^R=0.\label{lambday0R}
\end{eqnarray}
We can thus infer that $\zeta_x^R=\lambda_x^R+\lambda_{x0}^R$ and
$\zeta_y^R=\lambda_y^R+\lambda_{y0}^R$ are the complete solutions
for the classical equations of motion where $\lambda_x^R$ and $\lambda_y^R$ are
the particular solutions of the inhomogeneous equations
and $\lambda_{x0}^R$ and $\lambda_{y0}^R$ are the homogeneous
solutions baring the initial conditions.

In this manner, the center of the wave packet follows the same trajectory as
the classical particle.
This is a proof of Ehrenfest theorem. The trajectories
obtained for $\boldsymbol{\zeta}^R$ indeed are the same as the classical ones
as was proved by direct numerical calculations of the wave packet center motion shown
in Fig. (\ref{figure2}) with black crosses.
\section{Conclusions}\label{conclusions}

To summarize, we have studied the classical and quantum dissipation
of a charged particle in variable magnetic and electric fields through
a time dependent mass Hamiltonian.
To integrate the classical Hamiltonian, a series of three canonical transformations
are explicitly constructed and applied in order to reduce it to zero.
The final transformed variables
are at the same time constants of the motion and initial conditions
for the generalized momenta and positions. The final solution
to the equations of motion is rendered in its symplectic form.
Correspondingly, the quantum
Hamiltonian is reduced to zero by three unitary transformations. This procedure
allows for the calculation of the evolution operator in rather general
conditions, i.e. time dependent mass, variable electric and magnetic fields.
The generalized momentum and space variables in the Heisenberg picture
are expressed in terms of a
symplectic linear combination of their Schr\"odinger picture versions.
In times, the Green's function is constructed from the evolution operator
and the calculated expression is consistent with the structure
obtained by Schwinger and others \cite{schw1,schw2,schw3}.
As an example, the dynamics of a Gaussian wave packet under damping and
constant crossed
electric and magnetic fields is studied. Its motion is proved to follow
the same trajectory as the classical particle under the exact same
conditions.
The results presented in this paper might be useful in solid state
calculations where dissipation plays an important role.


We acknowledge support 
from UAM-A CBI projects 2232203, 2232204 and PROMEP project 2115/35621.
V.G. Ibarra-Sierra
acknowledges support from CONACyT.



\begin{thebibliography}{10}
\bibitem{comment1}%
  \BibitemOpen
  \bibinfo {note} {In order to test the physical sense of all the
  transformations up to now, we can consider the case of constant magnetic
  field and vanishing $\kappa$. In this case, the hamiltonian $H_2$ can be
  split in a time-independent Hamiltonian for two uncoupled harmonic
  oscillators and a smooth function of time. To eliminate the remaining time
  function, we introduce the additional coordinates $x_0=t$ and $p_0=H_2$ and
  the generating function
  $F_4=x_2p_{x4}+y_2p_{y4}+f\left(x_0\right)p_0+g\left(x_0\right)$. Therefore
  the final Hamiltonian is simply $H_4$ without any explicit time dependence.}%
  \bibAnnoteFile{Stop}{comment1}%
\end{thebibliography}
%
\end{document}